\documentstyle[11pt,epsf,bezier]{article}
\input{psfig}
\newcommand{\leeg}{\emptyset}
\newcommand{\A}{\mbox{A}}
\newcommand{\ceq}{c_{st}}
\begin{document}
\title{\bf{Real space renormalisation for reaction-diffusion systems}}
\author{Jef Hooyberghs\thanks{Aspirant Fonds voor Wetenschappelijk Onderzoek - Vlaanderen}\ , Carlo Vanderzande\\Departement WNI, Limburgs Universitair Centrum\\3590 Diepenbeek, Belgium\\
}
\maketitle

\ \\
\begin{abstract}
 The stationary state of stochastic processes such as reaction-diffusion systems can be related to the ground state
 of a suitably defined
quantum Hamiltonian. Using this analogy, we investigate the applicability of a real space 
renormalisation group approach, originally developped for quantum spin systems, to interacting particle systems.
We apply the technique to an exactly solvable reaction-diffusion system and to the contact process (both in
$d=1$). 
In the former case, several exact results are recovered. For the contact process, surprisingly good
estimates of critical parameters are obtained from a small-cell renormalisation. 
\end{abstract}
\newpage
\section{Introduction}
Reaction-diffusion systems and other interacting particle systems are relevant for the description
of several phenomena in physics,
chemistry and biology \cite{VK}. In the past, they have been mainly modelled by (non-linear) partial
differential equations \cite{MB}, a description which implicitly contains a mean-field assumption.
Such a description is however no longer appropriate in low dimensions where fluctuations are
important. To take these into account one turns to a description of the
reaction-diffusion system in terms of a stochastic process. This can for example be realised
by adding a noise term to the partial differential equation. In recent years however,
particular attention has been paid to models defined on a lattice. It has been found
that these can be related to a number of interesting topics in modern statistical mechanics
such as growing interfaces \cite{KPZ}, phase transitions into an absorbing state \cite{DR}, exactly solvable
quantum spin chains \cite{SH}, persistence exponents \cite{DerPer} and so on.

The systems which we will study here are defined on a lattice but evolve in continuous
time. At each site of the lattice, one can have (hard core) `particles' which can perform a random walk and/or
can undergo one or several `reactions'. In this paper we will limit ourselves to systems with one type
of particle . Each lattice site can then be either empty ($\leeg$) or be occupied by a particle ($\A$). 
As an example, consider a system
in which particles perform random walks and where two particles
on neighbouring sites can `annihilate' (i.e. undergo the reaction $\A+\A \to \leeg + \leeg$). 
In a simple mean field approach the density of
particles $c(t)$ in this system decays asymptotically as $1/t$. It is common to introduce a critical
 exponent $\theta$ which describes the decay of the density ($c(t) \sim t^{-\theta}$) and which therefore
in mean field theory equals $1$. An exact solution of the diffusion-annihilation model in $d=1$ (where it
is equivalent to the $T=0$ Glauber dynamics of an Ising model on the dual lattice) shows however that
$\theta=1/2$ \cite{AF}. This latter value is also found experimentally in systems which are thought to
be described by the diffusion-annihilation model. Moreover, $\theta$ shows a large universality across materials
and initial conditions. The same value for $\theta$ is also found when the experimental situation is described
more correctly by a diffusion-coagulation ($\A+\A \to \leeg + \A, \A+\A \to \A+\leeg$) model \cite{exp}. Hence, as in
the theory of equilibrium critical phenomena one needs a scheme which at the same time
explains the observed universality and gives precise values for critical exponents such as $\theta$. One is
therefore naturally led to a search for renormalisation group (RG) approaches to stochastic systems.

Mathematically, the stochastic process is a continuous time Markov process. Using the
formal equivalence between the master equation and the Schr\"{o}dinger equation in imaginary time, one can set up a
field theoretic formulation for stochastic systems which in turn can be used in the construction
of a renormalisation approach \cite{LC}. Critical exponents can then be calculated in
an $\epsilon$-expansion around the upper critical dimension, which in this case equals $2$. This approach
is by now well established and has been applied to many interesting systems.
As an example, we mention the much studied branching and annihilating random walks (BARW) \cite{BAW}, where besides
diffusion and annihilation, particles can undergo branching processes ($\A \to (m+1)\A$). The competition
between annihilation and branching leads to a non-equilibrium phase transition 
where the stationary state particle density $\ceq (\equiv \lim_{t \to \infty} c(t))$ goes
to zero in a continuous way as a function of the rates of the different processes. Near this transition, several critical exponents
(static and dynamic) can be introduced. It has been shown, both numerically and using the field theoretic RG, 
that the 
universality class of the transition for BARW is completely determined by the parity of
$m$ \cite{CT}. For $m$ odd, the transition falls in the universality class of directed percolation, whereas 
for $m$ even a new universality class appears. On the basis of very precise simulations, Jensen \cite{Jensen} conjectured
that $\theta=2/7$ and $\beta=13/14$ when $m$ is even ($\beta$ describes the way $\ceq$ goes to zero
near the transition). So far, a precise analytical calculation of
these exponents has however not been possible. Menyh\'{a}rd and \'{O}dor propose
the value $\beta=1$ on the basis of a perturbation around mean field theory \cite{MO}.  Using a loop expansion 
at fixed dimension, Cardy and T\"{a}uber
find $\beta=4/7$ \cite{CT}. Thus, there is clearly room for the introduction of new, analytical approaches
to the BARW and related models.

In this paper, we investigate the possibilities of a real space RG approach to interacting particle
systems. Our starting point is again the equivalence between a stochastic system and a quantum
mechanical one. In the past, several real space RG approaches to quantum lattice systems have been
introduced. We must mention here as an example the {\it density matrix renormalisation group} (DMRG)
which has been very succesful \cite{White1}. Currently, several research teams in the world are investigating
the applicability of the DMRG to reaction-diffusion systems \cite{CH},\cite{Pdmrg}. Unfortunately, the name DMRG
is a bit a misnomer, since one rarely calculates RG flows in this approach and hence it is not
easy to decide on questions of universality using this technique. Moreover, the approach is purely
numerical. Instead in this work, we will use an appropriate perturbative extension of 
the so called standard (also called SLAC) approach \cite{SLAC1}. We have applied this
RG method to the study of the stationary state properties of some interacting particle
systems. We found that the method
works surprisingly well in several cases which we studied. These include a solvable reaction diffusion
model (with diffusion, coagulation and decoagulation) and the contact process, which
is also thought to be in the directed percolation universality class. 

This paper is organised as follows. In section II we briefly introduce the description
of stochastic processes in a quantum mechanical language. In section III, we discuss a real space
approach to the ground state of quantum (spin or fermion) chains. Our
method is perturbative and in first order coincides with the standard (or SLAC) approach.
We discuss how the technique can be derived from more conventional (Niemeijer - Van Leeuwen)
RG approaches to quantum lattice systems by taking the zero
temperature limit. The method as such is
not new but, in our opinion, not very well known.
We also discuss how critical exponents for stochastic systems can be
obtained from such an approach. 
In section IV we  study properties of the stationary state of some solvable
reaction-diffusion systems. 
In section V, we give our results for the contact process. Finally,
in section VI we present some conclusions and an outlook on further
applications of the real space RG technique. 

\section{Quantum formalism of reaction-diffusion systems}

In this section we discuss the relation between stochastic processes in continuous time
on the one hand and quantum mechanics on the other hand \cite{Doi},\cite{Peliti}. This relation has been
(independently) discovered by several authors, and is now well established. We only give a brief
overview, with the main purpose of fixing our notation.

Consider a one dimensional lattice of $L$ sites and let $\eta=\{\eta_1,\ldots,\eta_L\}$ ($\eta_i=0(1)$ when no (a)
particle is present at site $i$) be the microscopic configuration of the particle system. Furthermore,
we denote by $P(\eta,t)$ the probability that the system is in configuration $\eta$ at time $t$.
The time evolution of $P(\eta,t)$ is determined by the transition rates $w_0,w_1,\ldots,w_s$ of the model.
In general one of these is used to fix the time scale (say $w_0=1$). We will collect the remaining
rates in a vector $\vec{w}=(w_1,\ldots,w_s)$. 

Next, we introduce a state vector $|P(t)\rangle$
\begin{eqnarray}
|P(t)\rangle = \sum_\eta P(\eta,t) |\eta\rangle
\end{eqnarray}
The time evolution of $|P(t)\rangle$ is described by the master equation
\begin{eqnarray}
\frac{d|P(t)\rangle}{dt}= - H |P(t)\rangle
\label{1.1}
\end{eqnarray}
where $H$ is a $(2^L \times 2^L)$ matrix which depends on all the transition rates of the system. In particular the matrix
element $-H_{\zeta,\xi} \geq 0$ (for $\zeta \neq \xi$) equals the transition rate to go from configuration $\zeta$ to configuration
$\xi$. Due to the conservation of probability the diagonal elements are given by
\begin{eqnarray}
H_{\zeta,\zeta}= - \sum_{\xi \neq \zeta} H_{\xi,\zeta} 
\label{1.2}
\end{eqnarray}
This relation implies that the sum of the elements in a column of $H$ equals zero, a condition
which we will refer to as the {\it stochasticity condition}. 

It is now common
to consider (\ref{1.1}) as a Schr\"{o}dinger equation (in imaginary time) and call $H$ the Hamiltonian.
In contrast to the quantum mechanical situation $H$ is not necessarily hermitian. Hence, left and right
eigenvectors are in generally not related by simple transposition.
In situations where there are only local interactions we can write $H$ in terms of local operators
$H=\sum_i H_i$. For reaction-diffusion systems with one type of particle, it is common to interpret the model
as a spin model (where a spin up (down) corresponds the absence (presence) of a particle) so that one
 writes $H$ in terms
of Pauli-matrices or
in terms of the matrices $E^{\alpha\beta}$ where $(E^{\alpha\beta})_{\sigma,\tau}=\delta_{\alpha\sigma}\delta_{\beta\tau}$.

The stochasticity condition implies the existence of at least one eigenvalue which equals zero and an associated
left eigenvector $\langle s|$ given by
\begin{eqnarray}
\langle s| =\sum_\eta \langle \eta|
\end{eqnarray}
Moreover, it can be shown that the real part of all the eigenvalues is non-negative.

The formal solution of the master equation (\ref{1.1}) is
\begin{eqnarray}
|P(t)\rangle = e^{-Ht} |P(0)\rangle
\end{eqnarray}
The expectation value of an operator, say $X$, corresponding with a physical quantity (such as the density
of particles) can then be written as
\begin{eqnarray}
\langle X(t) \rangle  &=& \sum_\eta X(\eta) P(\eta,t) \nonumber \\
&=& \langle s| X |P(t)\rangle \nonumber\\
&=& \langle s| X e^{-Ht} |P(0)\rangle
\label{1.3}
\end{eqnarray}
In this paper, we will be mainly interested in stationary state properties of the stochastic system.
In that limit, (\ref{1.3}) is determined by the groundstate(s) $|0\rangle$ of $H$. In particular,
(\ref{1.3}) becomes 
\begin{eqnarray}
X_{st}=\lim_{t\to\infty}\langle X(t)\rangle  = \langle s| X  |0\rangle
\label{1.3.1}
\end{eqnarray}
in the case of a non-degenerate ground state $|0\rangle$.

%Sometimes it can be convenient to make a {\it similarity transformation} which maps $H$ onto another
%Hamiltonian $\tilde{H}$ with the same spectrum. This can be achieved using an invertible matrix ${\cal B}$
%\begin{eqnarray}
%\tilde{H} = {\cal B} H {\cal B}^{-1}
%\end{eqnarray}

In summary, when we want to study properties in the stationary state of a stochastic system
we are in need of a (real space) RG which is suitable for the ground state of quantum spin systems.
In the next section, such an approach will be presented.

\section{Ground state renormalisation for quantum spin systems}
It is convenient to remind the reader briefly on the history of real space RG approaches to quantum systems.
In a first development, approaches working at finite temperature were introduced \cite{RD},\cite{ST}. These were
direct extensions of the first real space RG developped by Niemeijer and Van Leeuwen \cite{NVL}.
The SLAC approach was later introduced as an independent approach to the
ground state of quantum lattice systems. It was at first claimed to be non-perturbative. 
In a little known paper \cite{SVD}, it was however shown that the SLAC approach can be obtained from the
finite temperature technique by applying a suitable perturbation expansion (see below) to first order and by
sending the temperature $T \to 0$.
We now briefly describe the two methods and their relation. As will be explained below,
the possibility to turn the SLAC approach into a perturbative technique is of
importance for the study of some stochastic systems. 
 
The SLAC or standard approach was introduced in
\cite{SLAC1}. It was used a lot in the early eighties to study ground
state properties of several quantum spin and fermion chains (for a review, see \cite{PfJ}). As usual in real space RG
methods, the lattice is divided into cells, each containing $b$ sites. The Hamiltonian $H$
is then divided into an intracell part $H_{0}$ and an intercell part $V$. If $\alpha$ labels
the cells, one can in the particular case of one dimension write
\begin{eqnarray}
H = \sum_{\alpha} \left( H_{0,\alpha} + V_{\alpha,\alpha+1} \right)
\end{eqnarray}
As a first step, the Hamiltonian within one cell $H_{0,\alpha}$ is diagonalised exactly.
Let $H_{0,\alpha}$ have eigenvalues $E_{n,\alpha}$ with corresponding right (left)
eigenvectors $|n\rangle_\alpha (_\alpha\langle n|)$.  One then
selects two low lying eigenstates (for example the ground state $|0\rangle_\alpha$ and the
first excited state $|1\rangle_\alpha$) and considers them as effective spin states
for the cell: $|+_\alpha\rangle'=|0\rangle_\alpha$ and $|-_\alpha\rangle'=|1 \rangle_\alpha$.
Renormalised lattice states $|\sigma\rangle'$ 
can then be constructed by making tensor products over all cells: $|\sigma\rangle'=\otimes_{\alpha} |\sigma_\alpha\rangle'$.
These states span a $2^{L/b}$ dimensional vector space ${\cal W}$.

The renormalisation transformation, which always involves an elimination of degrees of freedom,
is now performed by projecting the original Hamiltonian onto ${\cal W}$. Mathematically this is achieved
by introducing
 a $2^L \times 2^{L/b}$-matrix $T_2$ whose columns contain the vectors $|\sigma\rangle'$ together
with a $2^{L/b} \times 2^L$-matrix $T_1$ whose rows contain the vectors $'\langle \sigma|$ (which
are constructed from the left eigenvectors of $H_{0,\alpha}$). Then the renormalised Hamiltonian $H'={\cal R}(H)$
is defined as
\begin{eqnarray}
H'={\cal R}(H) = T_1 H T_2
\label{2.1}
\end{eqnarray}
The transformation (\ref{2.1}) will define
a mapping in the parameter space $\vec{w}'=f(\vec{w})$, from which fixed points and critical
exponents can be determined as we will explain further below.

The procedure which was defined above at first sight seems to be rather {\it ad hoc} and non-perturbative.
Further insight in the method was obtained when its relation with more conventional real space RG
approaches was discovered \cite{SVD}. These approaches work at finite temperatures and are a direct extension of 
the Niemeijer-Van Leeuwen real space approach to the case of
quantum spin systems. In such an approach, the eigenstates of $H_{0,\alpha}$ are divided into two
groups according to some criterion (for example, a majority rule). Each group
is associated with an effective spin state for the cell. Within each group there
are $2^{b-1}$ states. Hence, cell states can be denoted as $|\sigma_\alpha,q_\alpha\rangle$, where
$\sigma_\alpha=\pm 1,\ q_\alpha=1,\ldots,2^{b-1}$. Tensor products of these states will be
denoted by $|\sigma,q\rangle=\otimes_{\alpha} |\sigma_\alpha,q_\alpha\rangle$. Finally
let ${\cal H}=-H/(kT)$ be the reduced Hamiltonian. The matrix elements of the renormalised (reduced) Hamiltonian
$\cal{H}'$ are then obtained by performing a partial trace
\begin{eqnarray}
'\langle \sigma|e^{{\cal H}'}|\sigma'\rangle'  = Tr_q\  \langle \sigma,q|e^{{\cal H}}|\sigma',q\rangle
\label{2.1.b}
\end{eqnarray}

Usually, it is impossible to work out this mapping exactly.
In any explicit calculation,
it is therefore necessary to perform an expansion in the intercell Hamiltonian $V$. This is achieved
by using the Feynman identity
\begin{eqnarray}
e^{\cal H}= e^{{\cal H}_0} \  T_{\lambda}\left[ \exp{\left( \int_{0}^{1} e^{-\lambda {\cal H}_0} {\cal V} e^{\lambda{\cal H}_0}d\lambda\right)}\right]
\label{2.2}
\end{eqnarray}
where $T_{\lambda}$ is a time ordering operator and ${\cal H}_0$ and ${\cal V}$ are respectively the
reduced intracell and intercell Hamiltonian. 
A similar expansion is made on the left hand side of (\ref{2.1.b}).
In reference \cite{SVD} it was shown that if one
expands (\ref{2.2}) {\it to first order} in $V$, and then takes the limit $T \to 0$, one recovers
the SLAC approach. This relation is useful for several reasons. First and most importantly, it shows that the SLAC
approach is perturbative and one obtains a procedure on how to calculate higher order corrections.
Secondly, since the finite temperature RG is constructed to conserve the partition function, one
is garanteed that the SLAC approach conserves the ground state energy. Finally, this relation
allows a consistent study of finite and zero temperature properties of a quantum
system in thermal equilibrium.

Going back to reaction-diffusion systems, which are non-equilibrium systems, we are thus garanteed
that our RG approach conserves the ground state energy. In principle, we also have a recipe
to calculate perturbative corrections to the SLAC approach. 
In this paper, we will limit ourselves to calculations in first order. Higher order corrections
usually lead to a proliferation of terms in the Hamiltonian.
However, as we will discuss in our conclusions, the existence of these higher order terms
will be necessary for a proper study of the BARW with $m=2$, or for any other model
in which $H$ contains terms involving three or more sites. 

While in this paper we will
only study stationary state properties (corresponding to ground state properties
of $H$), it is our feeling that the finite temperature extension of the SLAC-approach will
be useful to study finite $t$ properties of the reaction-diffusion systems.

We now turn to a discussion of how stationary state properties can be determined from
the RG mapping $\vec{w}'=f(\vec{w})$. 
To fix ideas, let us assume that this equation has a non-trivial fixed point at $\vec{w}=\vec{w^\star}$, 
with one relevant scaling field (which in linear approximation is proportional to
$\Delta w_1=w_1-w_1^\star$) whose scaling dimension is $y_{w_1}$. From standard
RG theory it then follows that near criticality, the correlation length $\xi$ will diverge
as $\xi \sim |\Delta w_1|^{\nu_\perp}$ with $\nu_\perp=1/y_{w_1}$.

In general it will be so that after renormalisation $w'_0 \neq 1$. Hence time needs to be
rescaled again, which is achieved by dividing $H'$ by $w'_0$. The number $w'_0(\vec{w}^\star)$
therefore teaches us how time rescales under a rescaling of space. We can use this quantity
to calculate the exponent $z$ as
\begin{eqnarray}
w'_0(\vec{w}^\star)=b^{-z}
\label{2.2.1}
\end{eqnarray}
From $\nu_\perp$ and $z$, scaling \cite{ScSP} gives us the exponent $\nu_\parallel=\nu_\perp/z$ which
determines the divergence of the relaxation time near the critical point. 
Next, we turn to the calculation of the particle
density $\ceq$. For a translationally invariant system, we can, using (\ref{1.3.1}), write $\ceq$
as
\begin{eqnarray*}
\ceq = \langle s| E^{11} |0\rangle 
\end{eqnarray*}
which under the RG transforms (\ref{2.1}) transforms as
\begin{eqnarray}
\ceq  &=&\langle s|T_1^{-1} T_1 E^{11} T_2 T_2^{-1}|0\rangle\\
&=& '\langle s| (E^{11})'|0\rangle'
\label{2.3}
\end{eqnarray}
The simplest possible case is the one in which the renormalisation of the
operator $E^{11}$ does not involve other operators so that $(E^{11})'=T_1 E^{11} T_2=a(\vec{w}) E^{11}$. In that
case (\ref{2.3}) becomes (where we now explicitly denote the dependence of $\ceq$ on the
transition rates)
\begin{eqnarray}
\ceq(\vec{w})= a(\vec{w}) \ceq(\vec{w}')
\label{2.4}
\end{eqnarray}
This relation can be iterated along the RG-flow, and hence the density of particles can be obtained
as an infinite product if one knows the density at the fixed point $\vec{w}_t^{\star}$ which attracts $\vec{w}$
(where the $t$ reflects the fact that this attractive fixed point is trivial and not critical).
\begin{eqnarray}
\ceq(\vec{w})= \left[\prod_{i=0}^{\infty} a(\vec{w}^{(i)})\right] \ceq(\vec{w}_t^{\star})
\label{2.4.b}
\end{eqnarray}
In principle correlation functions can be calculated in a similar way.

Near the critical fixed point $\vec{w}^{\star}$, we get from (\ref{2.4}) for the singular
part of $\ceq$ to leading order in $\Delta w_1$
\begin{eqnarray}
\ceq (\Delta w_1) = a(\vec{w}^{\star}) \ceq(b^{y_{w_1}}\Delta w_1)
\label{2.5}
\end{eqnarray}
From this relation we see that $a(\vec{w}^{\star})$ determines the rescaling of the particle
density at criticality. We write $a(\vec{w}^{\star})=b^{D-d}$ where $D$ can then
be interpreted as the fractal dimension of the sites that are occupied by particles at criticality.
Finally, from (\ref{2.5}) we get the behaviour of the particle density close to the critical point as
\begin{eqnarray}
\ceq (\Delta w_1) \sim \left(\Delta w_1\right)^{(d-D)/y_{w_1}}
\label{2.6}
\end{eqnarray}
so that we obtain the scaling relation $\beta=(d-D)\nu_\perp$.
Finally, $\theta$ can be obtained from the scaling relation $\theta=\beta/\nu_\parallel$ \cite{ScSP}.

It is straightforward to extend these reasonings to the case where the transformation of $E^{11}$ is more
complicated.

We thus see that a complete characterisation of the stationary state particle density and of all the
critical properties of the system can be obtained from our RG approach. 
In the next two sections we test our method on simple stochastic systems. The first one is
an exactly solvable reaction-diffusion system with a trivial transition. The second one
is the well known contact process which has a non trivial transition thought to be in the
DP universality class.

\section{Renormalisation for a simple reaction-diffusion system}
We consider a model with diffusion, decoagulation (which is the process $A+\leeg \to A+A$,\ $\leeg +A\to A+A$) and coagulation.
We will use the diffusion rate to fix the time scale, so that our model has two independent rates
which we will denote as $w_1$ (decoagulation) and $w_2$ (coagulation).  The local Hamiltonian
$H_i$ in this case can most conveniently be written as a $4\times 4$-matrix
\begin{eqnarray}
H_i = \left(\begin{array}{rrrr}
0 & 0 & 0 & 0\\
0 & 1+w_1 & -1 &-w_2\\
0 & -1 & 1+w_1 &-w_2\\
0 & -w_1 & -w_1 &2w_2\end{array}\right)   \label{3.4}
\end{eqnarray}
or in terms of the matrices $E^{\alpha\beta}$ as
\begin{eqnarray}
H_i &=& E^{11}_i E^{00}_{i+1} + E^{00}_i E^{11}_{i+1} + 2w_2(E^{11}_i E^{11}_{i+1}) + w_1(E^{11}_i E^{00}_{i+1} + E^{00}_i E^{11}_{i+1}) \nonumber \\
&-& E^{01}_i E^{10}_{i+1} - E^{10}_i E^{01}_{i+1} - w_2(E^{01}_i E^{11}_{i+1} + E^{11}_i E^{01}_{i+1}) -w_1(E^{11}_i E^{10}_{i+1} + E^{10}_i E^{11}_{i+1})  
\label{4.1}
\end{eqnarray}
With a similarity transformation this Hamiltonian can be mapped onto that of a free fermion system \cite{MSimL},
from which many exact results can be obtained.

To renormalise this model,
we take $b=2$, so that $H_{0,\alpha}$ is exactly given by (\ref{4.1}).
The spectrum of $H_i$ can be calculated trivially. The ground state is doubly degenerate and we identify
the corresponding eigenstates as effective spin states. In particular, we have (we use spin language
where $|+\rangle$ ($|-\rangle$) denotes a vacancy (an $A$-particle))
\begin{eqnarray}
|+\rangle' &=& |++\rangle  \nonumber\\
'\langle +|&=&\langle ++| \nonumber\\
|-\rangle'&=& \frac{1}{N} \left[ |+-\rangle + |-+\rangle + r|--\rangle \right] \nonumber \\
'\langle -|&=&\langle +-| + \langle -+| + \langle--|
\end{eqnarray}
where $r=w_1/w_2$ and $N=2+r$. 
Notice that we normalise the states as probability vectors (and not quantum mechanically).

Next, we calculate $H'$ using (\ref{2.1}). 
We find that the renormalised Hamiltonian contains the same terms as (\ref{4.1}). Moreover,
the transformation conserves the stochasticity condition. 
The renormalised diffusion rate is $1/N$. We divide the Hamiltonian by this factor (rescaling
of time). Then $H'$ has
completely the same form as $H$ but with renormalised values for $w_1$ and $w_2$.
The RG equations for $w_1$ and $w_2$
are
\begin{eqnarray}
w_1' = w_1 \frac{1+w_1+w_2}{w_2} \label{4.2.a}\\
w_2'= w_2 \frac{1+w_1+w_2}{2w_2 + w_1} \label{4.2.b}
\end{eqnarray}
The flow generated by these equations is shown in figure 1. There is a fixed point at $w_2=1,w_1=0$ (pure coagulation
fixed point). The line $w_2=1$ is
an invariant line. The RG equations, linearised at the fixed point, have the relevant eigenvalue $y_{w_1}=1$,
together with a marginal eigenvalue in the $w_2$ direction.

\begin{figure}
\setlength{\unitlength}{1cm}
\begin{center}
\begin{picture}(12,8)
\put(0,0){\vector(1,0){12}} \put(0,0){\vector(0,1){8}}
\put(-0.6,7.7){$w_1$}\put(11.7,-0.4){$w_2$}
\put(3.91,-0.07){$\bullet$}\put(4,-0.4){1}\put(4,0){\line(0,1){8}}\put(4,4){%
\vector(0,1){0.1}}
\put(2,0){\vector(1,0){0.1}}\put(6,0){\vector(-1,0){0.1}}\put(0,4){%
\vector(0,1){0.1}}
\bezier{500}(1,0)(3.8,0)(3.8,8)\put(3.56,4){\vector(1,4){0.01}}
\bezier{500}(0,0)(2,0)(2,8)\put(1.842,4){\vector(0,1){0.1}}
\bezier{500}(10,0.5)(4.2,0.5)(4.2,7)\put(10,0.49){\line(1,0){2}}\put(4.19,7){%
\line(0,1){1}}\put(4.62,4){\vector(-1,4){0.03}}
\bezier{500}(12,1.4)(5,2)(5,8)\put(6.17,4){\vector(-1,2){0.05}}
\end{picture}
\end{center}
\caption{Schematic RG-flow in the $(w_1,w_2)$-plane.}

\label{figure 1}
\end{figure}
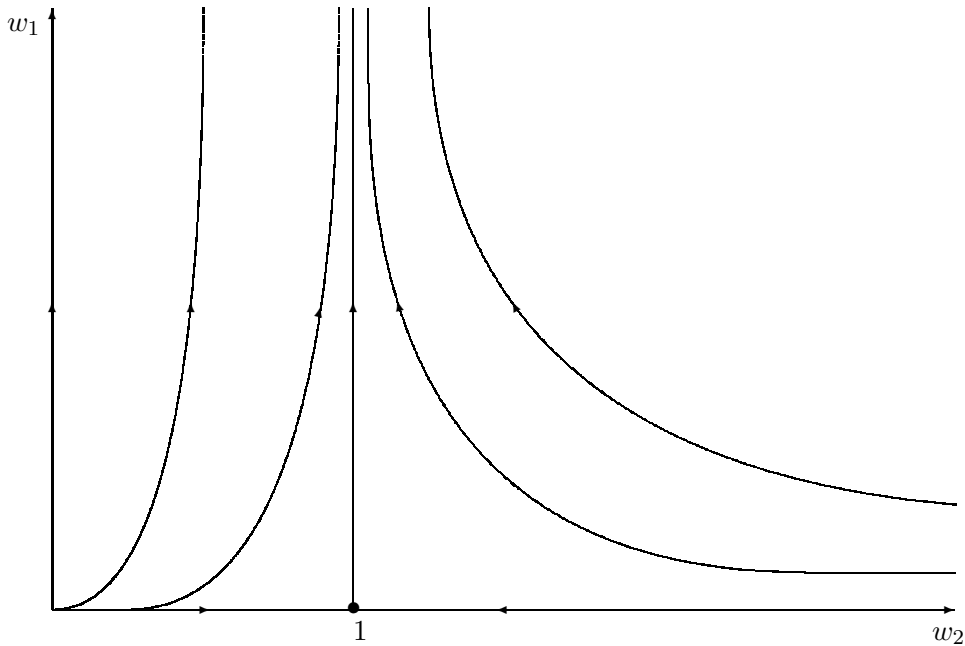

Next, we turn to a calculation of particle density in the stationary state using the scheme outlined
in the previous section. Projecting $E^{11}$ onto $\cal{W}$ we find the simple recursion
\begin{eqnarray}
(E^{11})' = \frac{1+r}{N} E^{11}
\label{4.2.c}
\end{eqnarray}
From this we have
\begin{eqnarray*}
a(w_1,w_2) = \frac{1+w_1/w_2}{2+w_1/w_2}
\end{eqnarray*}
Since also $w_1'/w_2'$ depends only on the ratio $w_1/w_2$ we arrive at the conclusion that
the particle density only depends on that ratio. Its precise value can then be calculated by making
a product of $a(w_1,w_2)$ along the RG flow. From figure 1 we see that all points with $w_1>0$ flow to points where
$w_1/w_2 \to \infty$. In that limit $\ceq=1$. Then, using (\ref{2.4.b}) we can obtain the
particle density. In figure 2 we plot the result for $\ceq$
as a function of $r$ together with the exact result \cite{MSimL}
\begin{eqnarray}
\ceq(r)=\frac{r}{r+1}
\label{2.exact}
\end{eqnarray}
Within the numerical accuracy, both results coincide. Hence it seems that we recover the exact result!
This might seem surprising since our calculation is only precise to first order in $V$.
With hindsight the accuracy of our result can be understood by the fact that the ground state of the whole system is
a product of one particle states. Nevertheless, this calculation illustrates nicely
the use of the method. Moreover it gives an RG-flow for the problem which cannot be obtained in any
other way.

\begin{figure}
  \centerline{ \psfig{figure=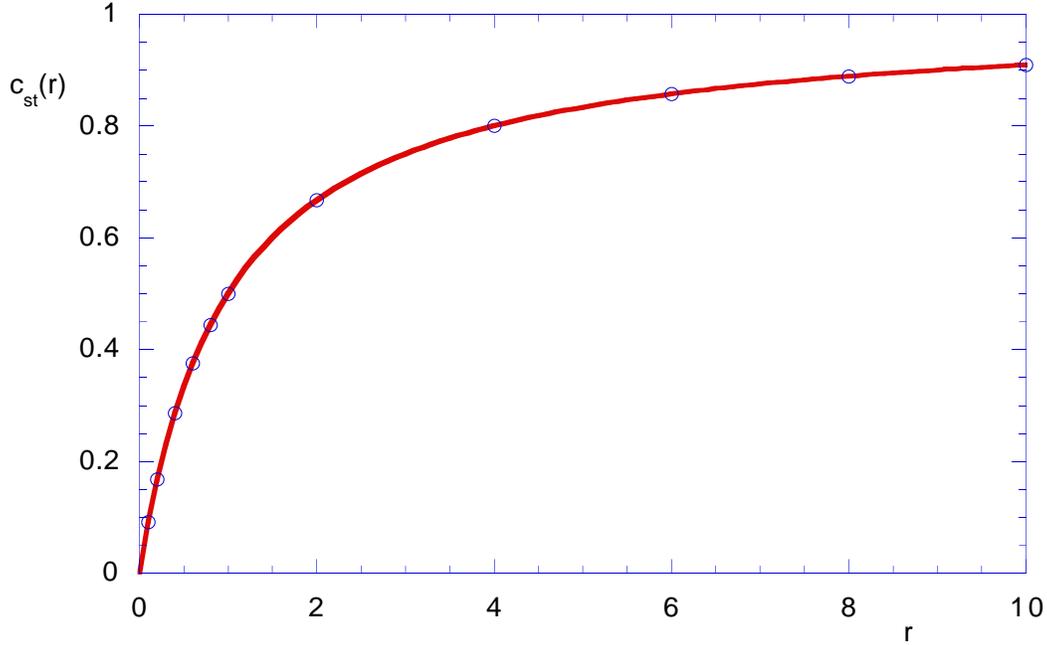,width=14cm}}

\caption{Particle density in the stationary state of the diffusion, coagulation, decoagulation 
 model. The full line represents the exact result, the  circles are the RG results.}

\label{figure 2}
\end{figure}

To conclude this section, we calculate the exponent $\beta$ which determines the behaviour of $\ceq$
near $r=0$. From the exact result (\ref{2.exact}), one obviously has $\beta=1$. On the other hand, from
 (\ref{4.2.c}), we have
$a(0,1)=1/2$, hence $D=0$ and using $y_{w_1}=1$ we recover the exact result $\beta=1$.

At this place it is appropriate to mention that for other simple reaction diffusion systems 
we can also recover exactly known results. An example is a model with diffusion, decoagulation
and death (which is the process $A+\leeg\to\leeg+\leeg,\ \leeg+A\to\leeg+\leeg$). This model undergoes a first order
transition \cite{MSimL}. When the decoagulation rate is greater then the death rate, $\ceq=1$, whereas
when the opposite inequality holds, one has $\ceq=0$. The RG recovers this exact result.
As usual the first order transition is controlled by a discontinuity fixed point.

Also for a model with diffusion, annihilation and pair creation $\leeg+\leeg \to A+A$, we recover
the exact result, first obtained by Z. R\'{a}cz \cite{Racz}, that $\ceq$ grows with the square root of
the pair creation rate.

\section{The contact proces}
The contact proces was originally introduced as a simple model for an epidemic \cite{Harris}. In that
interpretation a particle corresponds to sick person, and a vacancy to a healthy individual.
In the process particles can disappear ($A\to 0$) with a rate $w_0=1$. Empty sites can
become occupied with a rate $\lambda z/2$, where $z$ is the number of occupied neighbours
of the empty site (this represents contamination in epidemic terms).
This model cannot be solved exactly, but its critical exponents 
are known to high accuracy \cite{DR}. On the basis of numerical data, and from symmetry arguments, it is
generally believed that the model is in the DP-universality class \cite{DPJ},\cite{DPG}.

The quantum Hamiltonian corresponding to this model contains one-site terms (for
the process $A\to 0$) and two-site terms (for the contamination process).
The local Hamiltonian for the model is
\begin{eqnarray}
H_i = \left(\begin{array}{rrrr}
0 & 0 & -1 & 0\\
0 & \lambda/2 & 0 & -1\\
0 & 0 & 1+\lambda/2 & 0\\
0 & -\lambda/2 & -\lambda/2 &1\end{array}\right)   \label{5.1}
\end{eqnarray}
We have split the whole Hamiltonian in such a way that $H_i$ contains
one two-site contribution and one one-site term. In this way, when performing
the RG , both $H_{0,\alpha}$ en $V_{\alpha,\alpha+1}$ will contain an equal
number of one-site and two-site terms. This garantees, at least for
the contact process, that the ground state  of the intracell Hamiltonian is a doublet.
These states are then the natural candidates to be used as effective cell spins.

We will perform a renormalisation for this model using a cell with $b=3$. The calculation
is straightforward and can most effectively be done using $^\copyright$Mathematica.
It came as a surprise to us that also in this case, there is no proliferation
of interactions in the renormalised Hamiltonian. The RG-equation for $\lambda$ is
\begin{eqnarray}
\lambda'=\frac{\lambda^3 (2+\lambda) (8+10\lambda+4\lambda^2+\lambda^3)}{4(16+40 \lambda+37 \lambda^2
+18\lambda^3+4\lambda^4)}
\label{5.2}
\end{eqnarray}
This equation has one non-trivial repulsive fixed point at $\lambda=\lambda^\star=3.22319$.
This value is surprisingly close to the best known numerical value for the contact process
which is $\lambda_c=3.2978$ (all numerical results are taken from \cite{JDS}). 
From linearising the RG-equations near the fixed point,
we obtain $y_{w_1}=.8886$, from which we obtain $\nu_\perp=1.1253$, to be compared
with the numerically determined value of $1.0972$.
As explained in section 3, one can obtain the exponent $z$ from the rescaling of
the unit of time. In this case we have
\begin{eqnarray}
w_0'=\frac{4(16+40\lambda+37 \lambda^2+18\lambda^3+4\lambda^4)}{(8 +10\lambda+4\lambda^2+\lambda^3)^2}
\label{5.2.b}
\end{eqnarray}
from which we obtain $z=.6858$ (to be compared with $z=.636$).
Finally we need to determine $D$. As explained in section 3, one therefore has to renormalise
the operator $E^{11}$. In a cell with $b \geq 3$ this can be done in several ways. Either
one projects the operator $E^{11}_m$ where $m$ is a site at or near the middle of the cell.
Alternatively we can take an `average' operator $E^{11}_a$
\begin{eqnarray*}
E^{11}_a=\frac{1}{b}\sum_{i=1}^b E^{11}_i
\end{eqnarray*}
We have performed both calculations. They give respectively $D=.6391$ and $D=.6940$, from
which we obtain $\beta=.4061$, respectively $\beta=.3444$. These values should be compared
with the precise numerical estimate $\beta=.2769$. Other exponents can be obtained
using scaling relations. We get $\nu_\parallel=1.6409, \theta=.2099$ whereas the best
known values are $\nu_\parallel=1.736, \theta=.1597$.
We thus see that, taking into account the smallness of the cell considered,
our estimates of $\lambda_c$ and all critical exponents are
close to the known values.

In figure 3 we finally plot $\ceq(\lambda)$ as obtained from our RG approach. At the
scale of the figure it almost coincides with the results from other approaches.

\begin{figure}
  \centerline{ \psfig{figure=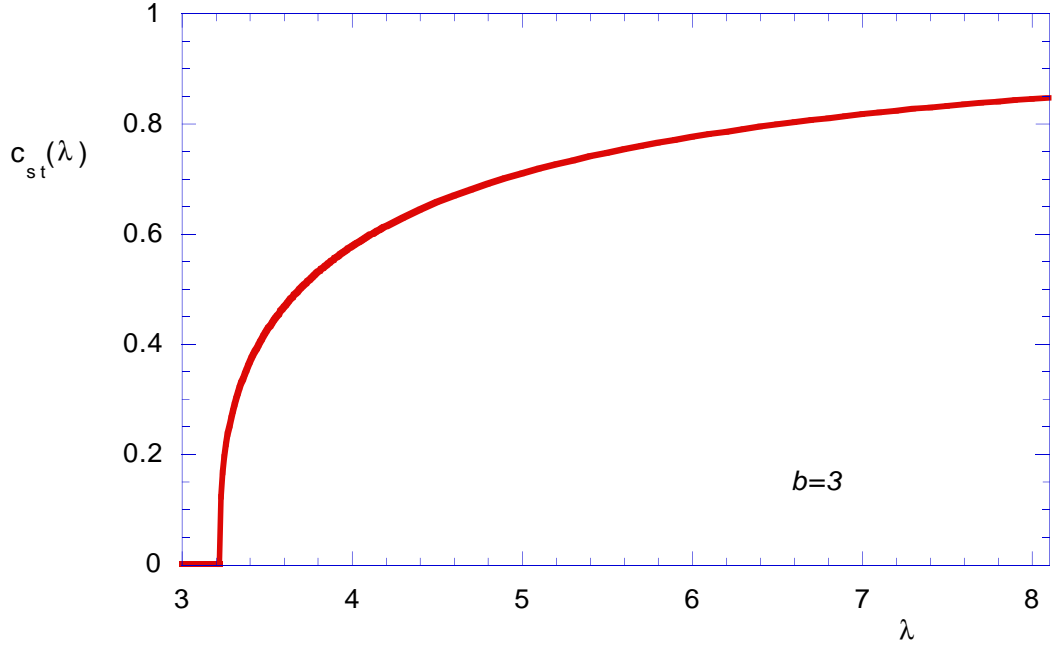,width=14cm}}

\caption{Particle density in the stationary state of the contact process as obtained from the RG approach.}

\label{figure 3}
\end{figure}

We are currently extending our calculations for the contact process to larger cell sizes. We hope that
our approach, combined with suitable extrapolation techniques, is able to
give very precise estimates of critical exponents.

\section{Conclusions}
In this paper we have investigated the applicability to reaction-diffusion
processes of a quantum real space RG method. We first studied simple processes for
which exact information is available. In three cases, these exact results are reproduced
in calculations done on very small cells. 

So far, we encountered one system in which the RG predicted wrong results.
This is a model with diffusion, coagulation and birth ($\leeg+\leeg\to\leeg+A,\ 
\leeg+\leeg\to A+\leeg$). When the diffusion rate equals the coagulation rate, this
model belongs to a class which is, at least partially, integrable \cite{PeshelI}. For small birth rates
$w_3$, $\ceq$ is known to grow as $w_3^{1/3}$. The application of our technique to that
model raised several difficulties. The ground state of $H_{0,\alpha}$ is a singlet, and for the
first excited state there is a crossing of energy levels (in finite systems).
Hence, there is no obvious choice for the effective cell states. 
For a specific choice we made, the renormalised Hamiltonian $H'$ turns out to have
a form different from the original Hamiltonian $H$, but there is a similarity
mapping this $H'$ onto $H$ with renormalised couplings, and an extra decoagulation term. If we define the
full RG as the projection (\ref{2.1}), followed by the similarity we are
able to obtain a flow in parameter space. Unfortunately, our
results indicate that for cells with $b=2$ and $b=3$, $\ceq$ grows as $w_3^{1/2}$. 
We hope to clarify the RG for this model in the future.

For the contact process, which contains a non-trivial critical point,
rather accurate estimates for the location of the critical point and
for critical exponents are obtained from a calculation on a cell of $3$ sites.
This justifies the hope that by going to larger cells and using good
extrapolation techniques very precise exponent estimates can be obtained.
We are currently performing such calculations.

Another project we hope to carry through is a study of the BARW with $m=2$. As stated in
the introduction, there are few reliable analytical results on the critical properties of this system. 
A particularly nice model that is known to be in this universality class was introduced
by Menyh\'{a}rd. It is a non-equilibrium Ising model (NEKIM) with Glauber dynamics at zero
temperature and Kawasaki dynamics at infinite temperature \cite{NEKIM}. It was recently shown that
this model is selfdual \cite{MuS}. The ground state energy of the quantum Hamiltonian corresponding
with this model is for all finite systems again doubly degenerated,
so that effective cell states can be defined unambiguously.
The quantum Hamiltonian for the NEKIM contains a three-site interaction
term. Under the RG, performed to first order, such a term will be mapped onto a two-cell interaction term.
However, by extending the SLAC-approach to second order in $V$, as 
discussed in section 3, one could generate
a renormalised three-cell interaction. 
In fact, from a mathematical point of view, the Hamiltonian of the NEKIM is rather similar
to that of a transverse Ising model with three spin interactions \cite{TIM31},\cite{TIM32}. That model
also has a selfduality and was renormalised successfully by using our approach to second
order \cite{CVD}. 
An alternative model in the universality class of the even-$m$-BARWis a recently introduced variant
of the contact process in which particles disappear and are contaminated in pairs \cite{NCP}.
Since this model doesn't involve any diffusion, it may be more simple to analyse.

It has of course to be admitted that real space RG methods in general involve some ill
understood approximations. Nevertheless, it is our opinion that the results
presented here give considerable hope that our RG can succesfully be used to further understand
the critical behaviour of non
exactly solved systems such as the BARW.

{\bf Acknowledgement} We thank M. Henkel, P. Leoni and G. M. Sch\"{u}tz for some interesting discussions on
the subject of this paper. We also thank the IUAP for financial support. One of us (JH)
thanks the ``Fonds voor Wetenschappelijk Onderzoek - Vlaanderen'' for financial support.

\end{document}